\documentclass[aps,twocolumn,prl,floatfix,showpacs]{revtex4}
\usepackage{graphicx}
\usepackage{longtable}
\begin{document}
\pagestyle{plain}

\title{ Enhanced heating of salty water and ice under microwaves: 
Molecular dynamics study}

\author{Motohiko Tanaka, and Motoyasu Sato}

\affiliation{Coordination Research Center, National Institute for 
Fusion Science \\
Oroshi-cho, Toki 509-5292, Japan; Email: mtanaka@nifs.ac.jp}

\begin{abstract}
By molecular dynamics simulations, we have studied the enhanced 
heating process of salty ice and water by the electric field of 
applied microwaves at 2.5 GHz, and those in the range 2.5-10 GHz 
for the frequency dependence. 
We show that water molecules in salty ice are allowed to rotate 
in response to the microwave electric field to the extent 
comparable to those in pure water, because the molecules 
in salty ice are loosely tied by hydrogen bonds with adjacent 
molecules unlike the case of rigidly bonded pure ice. 
The weakening of hydrogen-bonded network of molecules in salty 
ice is mainly caused by the electrostatic effect of salt ions 
rather than the short-range geometrical (atom size) effect of 
salt ions since the presence of salt ions with small radii 
results in similar enhanced heating.
\end{abstract}

\maketitle

\begin{flushleft}
{\bf 1. Introduction}
\end{flushleft}

Microwaves are frequently utilized as a highly efficient and less 
carbon-oxide releasing energy source for cooking or drying food, 
sintering metals and ceramics. 
It is known that dielectric materials including pure and salty water 
are heated efficiently by absorbing the electric field energy of 
microwaves [von Hippel 1954; Hobbs et al., 1966; Hasted, 1972; 
Buchner, 1999; Meissner and Wentz, 2004; Takei, 2007]. 
It is also known that metallic powders and magnetic materials 
such as magnetite and titanium oxides with oxygen defects 
TiO$_{2-x} $ ($ x > 0 $) are heated by the magnetic field of 
microwaves [Roy et al, 1999; Peelamedu et al. 2002; 
Sato et al. 2006; Suzuki et al. 2008; Tanaka et al., 2008].
   
   In order to examine the mechanism of microwave heating of water 
and ice at the molecular level, we performed molecular dynamics 
simulations [Tanaka and Sato, 2007]. 
There we showed mechanistically that (i) rotational excitation of 
electric dipoles of water coupled with irreversible energy relaxation 
to the translational energy was responsible for the microwave heating 
of water, (ii) pure ice was not heated by microwaves because of 
strong hydrogen bonds between water molecules, (iii) the experimentally 
known enhanced heating of salt water by microwaves was due to 
acceleration of salt ions by the microwave electric field, 
which is usually termed as Joule heating. 
We numerically evaluated the time integrated Joule heating term of 
salt ions $ \bf{E} \cdot \bf{J} $ and the work associated with 
rotation of electric dipoles $ \bf{E} \cdot d\bf{P}/dt $ by summing up 
the contribution of all atoms or molecules, where $ \bf{E} $ is 
the electric field of microwaves, $ \bf{J}= \sum_{i} q_{i} \bf{v}_{i} $ 
is the current carried by Na$^{+} $ and Cl$^{-} $ ions, 
and $ \bf{P}= \sum_{s} \bf{d}_{s} $ is the sum of electric dipoles of 
all water molecules, with $ q_{i} $ and $ \bf{v}_{i} $ the charge 
and velocity of i-th atom, respectively, and $ \bf{d}_{s} $ 
the electric dipole moment of s-th molecule. 
We obtained the relation 
$ |\bf{E} \cdot \bf{J}| \cong {\rm 2} |\bf{E} \cdot d\bf{P}/dt| $ 
for dilute saline solution of 1 mol\% salinity and 10 GHz microwave 
[Tanaka and Sato, 2007].  
   
     We stated above that pure ice is not heated by microwaves. 
However, we have a daily experience of processing (melting) frozen food 
in microwave ovens.  
Our motivation of the present work is to examine the heating of salty ice 
as well as salty water by molecular dynamic simulation. 
In this paper we mainly use the microwaves with 2.5 GHz, except for 
5-10 GHz microwaves for examining the frequency dependence of the heating.

\begin{flushleft}
{\bf 2. Simulation Procedures}
\end{flushleft}

Ice and liquid water are characterized by a completely and partially 
ordered network of H$_{2}$O molecules, respectively, mediated by hydrogen bonds 
connecting adjacent two molecules [Matsumoto et al., 2002]. 
For this reason, we adopt an explicit water model with three point charges 
- the SPC/E water model [Toukan and Rahman, 1985; Berendsen et al., 1987]. 
A typical simulation run requires the computation time of a few days 
for roughly 3000 water molecules on our Opteron-based cluster machine
(usually four processors are used per job). 
Although the model has reasonable electrostatic properties, 
for more accurate calculation of the energy absorption, the use of 
TIP4P-FQ model with polarization effect [English and MacElroy, 
2003; English, 2006] may be preferred but with small number of water 
molecules due to large computation times.

   We follow the same numerical procedures as those used previously 
[Tanaka and Sato 2007]. We start all the simulations with the crystal ice 
and relax it to the desired temperature before microwaves are applied. 
This is because molecules of initially random orientations take 
a very long time to arrive at the equilibrium with properly ordered 
hydrogen-bonded network. 
A water molecule of the SPC/E model in Fig.1 illustrates that three atoms 
form a rigid triangle H-O-H with partial charges $ q (=0.424e) $ and 
$ -2q $ residing at the hydrogen and oxygen sites, respectively. 
The electrostatic forces are calculated for pairs of these atoms 
while the Lennard-Jones forces are calculated only for the pairs 
of oxygen atoms. 
The equations of motion are
\begin{eqnarray}
\hspace*{-0.2cm} && m_{i} \frac{d \bf{v}_{i}}{dt} =
-\nabla \left[ \sum_{j} \frac{q_{i}q_{j}}{r_{ij}} 
 +48 \varepsilon_{ij} \left\{ \left( \frac{\sigma_{ij}}{r_{ij}} \right)^{12} 
                             -\left( \frac{\sigma_{ij}}{r_{ij}} \right)^{6} 
                      \right\} \right]
 \nonumber \\
\hspace*{-0.2cm} && \hspace*{1.5cm} + q_{i} \bf{E}_{mw} \\
\hspace*{-0.2cm} && \frac{d \bf{r}_{i}}{dt} = \bf{v}_{i}
\end{eqnarray}
Here, $ \bf{r}_{i} $ and $ \bf{v}_{i} $ are the position and velocity 
of i-th atom, respectively, $ m_{i} $ and $ q_{i} $ are its mass and 
charge, respectively, and $ r_{ij}= |\bf{r}_{i} - \bf{r}_{j}| $
is the distance between i-th and j-th atoms. 
The second term in the bracket on the right-hand side of Eq.(2) serves 
for two atoms not to overlap each other, where $ \sigma_{ij} $ 
is the sum of radii of two atoms and $ \varepsilon_{ij} $ is 
the Lennard-Jones energy. 
For water, $ \varepsilon_{ww}= 0.65 $ kJ/mol and $ \sigma_{ww}= 0.317 $ nm. 
Salt ions are put randomly in space avoiding water molecules and 
are treated as isolated species with the charge $ q_{Na}= e $, 
the diameter $ 2 \sigma_{Na}= 0.26 $ nm, and the Lennard-Jones energy 
$ \varepsilon_{Na}= 0.062 $ kJ/mol for Na$^{+} $ and $ q_{Cl}= -e $, 
$ 2 \sigma_{Cl}= 0.44 $ nm and $ \varepsilon_{Cl}= 0.446 $ kJ/mol 
for Cl$^{-} $ [AMBER, 2008]; the hybrid rule
\begin{eqnarray}
\varepsilon_{ij} = (\varepsilon_{i} \varepsilon_{j})^{1/2}, \ \ 
\sigma_{ij}=  \sigma_{i} + \sigma_{j}
\end{eqnarray}
is used when two species $ i $ and $ j $ are encountering. 
The diameter of Cl ion is comparable with the size of the network cell 
whose building blocks are six-membered water rings. 
The {\small SHAKE} and {\small RATTLE} algorithm [Andersen, 1983] is adopted 
in time integration of Eq.(1) and (2) to keep the rigid triangular 
shape of the molecule. The time step for integration is 1 fs. 

\begin{figure}
\centerline{\scalebox{0.50}{\includegraphics{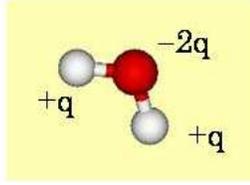}}}

\vspace*{-0.25cm}
\caption{
Three-atom rigid-body model for a water molecule (SPC/E  model) with 
partial charges on atom sites.
}
\label{Fig.1}
\end{figure}

   Our simulation system consists of a cubic box of each side 4.2nm 
that contains (14)$^{3}$ water molecules when salt is not added. 
The periodic boundary condition is imposed under the constant 
system volume. 
Thus, the Ewald sum for infinite number of periodic image charges 
is taken to correctly account for the electric field in which 
the particle-particle particle-mesh algorithm is adopted 
[Deserno and Holm, 1998]. 
The real space cutoff distance for both the electrostatic and 
Lennard-Jones forces is 1.0 nm, and 32 spatial grids are used for 
the Fourier-space calculation of the nonlocal electrostatic forces. 
   
   After an equilibration phase of 100 ps, a spatially homogeneous 
microwave electric field of the form 
$ \bf{E}_{mw}({\it t})= E_{0} \sin \omega {\it t} \ \hat{\bf{x}} $  
is applied. 
The wave frequency $ f= \omega /2\pi $ is 2.5 GHz unless otherwise 
specified. 
To overcome the numerical noises in the simulation, the used wave 
amplitude is $ E_{0}= 2.3 \times 10^{6} $ V/cm which is large 
but its associated energy with the electric dipole of a water 
molecule is less than the thermal energy at room temperature 
$ E_{0}d/k_{B}T_{300K} \cong 0.4 $. 
This condition assures us the safe use of this electric field 
in our simulations, where $ d \cong 7.8 \times 10^{-30} $ Cm  
is the dipole moment of the model water molecule.

\begin{flushleft}
{\bf 3. Results and Discussion}
\end{flushleft}

The initial condition of the simulation is prepared by generating 
crystal ice of the I$_{c} $ structure,  and putting Na and Cl ions 
of 1 mol\% concentration (roughly 3 wt\%) when salty ice or water 
is prepared. 
The I$_{c} $ ice is isotropic in three directions whose physical 
properties are quite similar to those of the ordinary ice of the 
I$_{h} $ structure [Eisenberg and Kauzman, 1969; 
Matsumoto et al., 2002]; 
we swap the O-H bonds of water molecules successively until an 
isotropic state has been reached. 
The initial structure is equilibrated for 100 ps at a given temperature, 
and then the microwave electric field of 2.5 GHz frequency is applied. 

\begin{figure}
\centerline{\scalebox{0.75}{\includegraphics{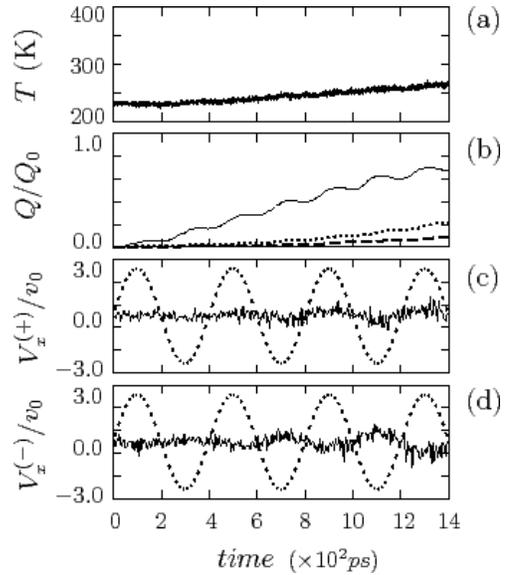}}}

\vspace*{-0.25cm}
\caption{
The time history of the heating of salty ice, (a) temperature of the 
whole system in Kelvin, (b) the dipole heating term (solid line), 
the Joule heating term of all salt (dotted) and that of Na ion (dashed), 
(c) the x-component velocity of Na ions, and (d) that of Cl ions with 
$ v_{0}= 10 $ m/s, $ Q_{0}= 10^{-9} $ J. The amplitude of the 
microwave electric field is shown by dotted lines in (c) and (d), 
where the electric field of 2.5 GHz is turned on just after the 
equilibration phase at time 0 of the figure. 
}
\label{Fig.2}
\end{figure}

Figure 2 shows the time history of (a) the temperature (starting at 
230 K) of the salty ice system, (b) time integrated value of the 
Joule heating and dipole heating terms, (c) the average velocity of 
Na$^{+} $ along the x direction, and (d) that of Cl$^{-} $. 
It is noted that the temperature of salty ice increases monotonically 
after the microwave is turned on. 
The velocities of salt ions in panels (c) and (d) remain small 
compared to those in salty water in Fig. 3, until the ice approaches 
the melting temperature. 
To clarify the heating mechanism, the Joule heating term 
$ \bf{E} \cdot \bf{J} $ (dotted and dashed lines for the total salt 
and Na$^{+} $, respectively) and the dipole heating term 
$ \bf{E} \cdot d\bf{P}/dt $ (solid line) are plotted separately in (b). 
Evidently, the excitation of dipole rotation is the principal heating 
mechanism of salty ice. 

\begin{figure}
\centerline{\scalebox{0.75}{\includegraphics{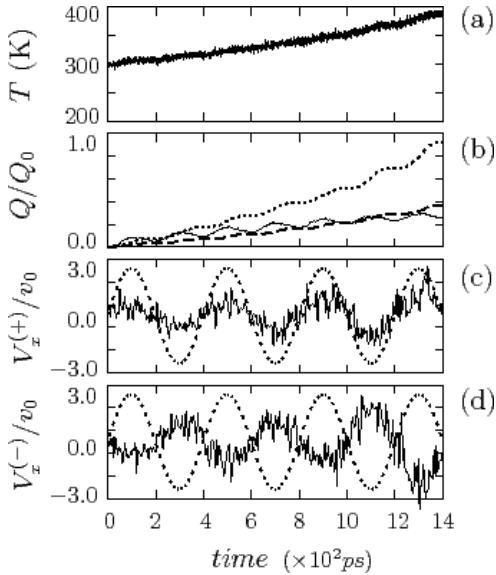}}}

\vspace*{-0.25cm}
\caption{
The time history of the heating of salty water, (a) temperature of 
the whole system in Kelvin, (b) the dipole heating term (solid line), 
the Joule heating term of all salt (dotted) and that of Na ion (dashed), 
(c) the x-component velocity of Na ions, and (d) that of Cl ions with 
$ v_{0}= 10 $ m/s, $ Q_{0}= 10^{-9} $ J. The amplitude of the 
microwave electric field is shown by dotted lines in (c) and (d), 
where the electric field is turned on just after the equilibration 
phase at time 0 of the figure. 
}
\label{Fig.3}
\end{figure}

    The heating of salty ice is compared with that of salty water in Fig.3. 
Salty water with 1 mol\% salinity is prepared in the same way as for 
salty ice, and is melted and equilibrated for 100 ps at 300 K. 
For the salty water, the heating rate increases with temperature above 350 K, 
as seen in Fig.3(a) at which the slope of the curve becomes steeper. 
The velocity oscillations of Na and Cl ions in (c) and (d) also increase 
in time and are much larger than those for salty ice. 
The large oscillations at later times reveal de-trapping of salt ions 
from the cell of water network, which accelerates the heating due to 
the large amplitude of oscillations. 
A striking difference is found in the heating terms in panel (b): 
the Joule heating term much exceeds the dipole heating term, 
where the former is twice larger than the latter as shown previously 
[Tanaka and Sato, 2007]. 
It is interesting that Cl ions with heavier mass and larger radius 
respond more than Na ions to the microwave electric field. 
This implies that Cl ions are not stably contained in the cells of 
the hydrogen-bonded network of water molecules, which makes them 
more easily be de-trapped and move through the network.     

   As for reference, the heating of pure water against 2.5 GHz microwave 
is shown in Fig. 4. 
In this case, the excitation of dipole rotation is the only mechanism of 
energy absorption from the microwave. 
The heating rate of salty ice with 1 mol\% salinity is comparable to that 
for pure water. 
The velocity oscillations are shown for (c) hydrogen atoms and (d) oxygen 
atoms, which are small compared with those in Fig. 3 and are out of phase 
with each other since the center of mass of water molecules is fixed 
because of the charge neutrality.

\begin{figure}
\centerline{\scalebox{0.75}{\includegraphics{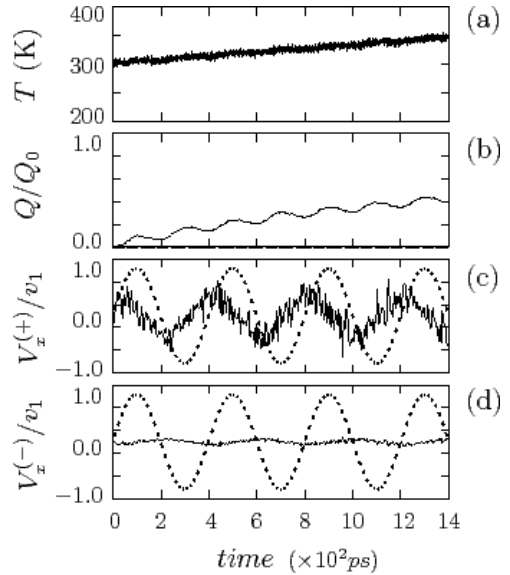}}}

\vspace*{-0.25cm}
\caption{
The time history of the heating of pure water, (a) temperature in Kelvin, 
(b) the dipole heating term, (c) the x-component velocity of hydrogen atoms, 
and (d) that of oxygen atoms with $ v_{1}= 1 $ m/s, $ Q_{0}= 10^{-9} $ J. 
The amplitude of the microwave electric field is shown by dotted lines 
in (c) and (d), where the electric field of 2.5 GHz is turned on just after 
the equilibration phase at time 0 of the figure. 
}
\label{Fig.4}
\end{figure}
   
     As pure ice is not heated by microwaves of the GHz band, 
the presence of salt in ice should be playing a decisive role in the heating 
(melting) process in the microwave electric field. 
We showed on the energy basis in Fig. 2 that not the Joule heating but 
the excitation of dipole rotation is responsible for the heating. 
Then it follows that 
{\it electric dipoles of water molecules are allowed to rotate in salty ice,} 
which occurs because of the local breaking of water network by salt ions. 
This is well recognized in Fig. 5 which shows the time history of the sum 
of the electric dipoles for (a) pure water and (b) salty ice. 
For pure water, molecules begin rotational motions immediately with the 
microwave electric field. 
In the early phase when microwaves are applied to salty ice, the rotational 
motion of the dipoles begins which is somewhat retarded and small 
compared to the case of pure water. This is completely different from 
the case of pure ice. 

\begin{figure}
\centerline{\scalebox{0.75}{\includegraphics{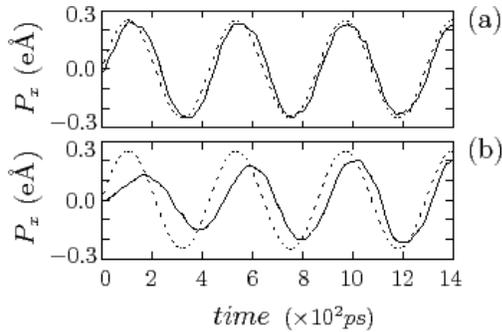}}}

\vspace*{-0.25cm}
\caption{
Time history of the sum of all electric dipoles in the x direction for 
(a) pure water (corresponding to Fig.4) and (b) salty ice (Fig.2). 
Solid lines show the value of electric dipoles, and dashed lines show 
the amplitude of the electric field of microwaves. 
}
\label{Fig.5}
\end{figure}

   Figure 6 shows that the rigid network of ice is distorted quickly 
in the presence of salt ions even before microwave is applied, as seen in (b). 
Experimentally, anomalously high permittivity was reported for salty ice 
[Zhong et al, 1988]. 
Here, it is confirmed by numerical simulation at the molecular level 
that the high response of salty ice to microwaves is caused by the 
presence of salt ions in ice which weaken the hydrogen-bonded rigid network 
of water molecules.

\begin{figure}
\centerline{\scalebox{0.41}{\includegraphics{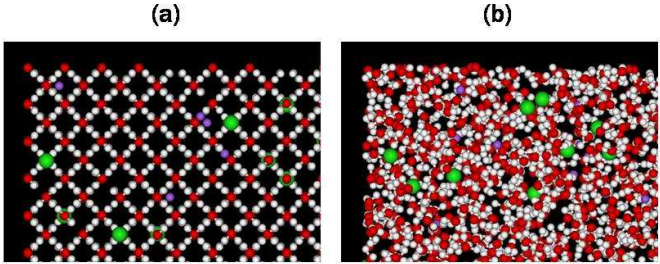}}}

\vspace*{-0.05cm}
\caption{
3D view of water molecules and salt ions for salty ice 
(230 K) at (a)  the initial stage, and  (b) after 100 ps without microwave 
application. Chained gray and white spheres represent oxygen and hydrogen atoms, 
respectively, Na and Cl ions are shown by isolated small and large sphere, respectively. 
}
\label{Fig.6}
\end{figure}
   
   The weakening of the hydrogen-bonded network of water molecules 
which occurs in salty ice is also confirmed by specially designed simulation: 
the diameter of charged salt ions is artificially reduced to 0.2 nm so that 
they nicely fit in the cell of the network. Even in such a 
condition, heating of salty ice does take place whose heating rate is 
almost three quarters that of the normal case. This proves that the weakening of 
the network in salty ice is caused mainly by the electrostatic effect of salt ions 
which is long-ranged, rather than the short-range geometrical effect.

     The dependence of the heating rate on the frequency of microwave is 
examined for pure and salty water and salty ice, as depicted in Fig. 7. 
Microwaves of three frequencies 2.5, 5 and 10 GHz are used, and the temperature 
increase in the 1.4 ns interval is plotted for pure water (filled circles), 
salty water (open circles) and salty ice (triangles). 
The salinity of salty ice and water is 1 mol\%, and initial temperature is 300 K 
for water and 230 K for ice. The line fitting yields the frequency dependence of 
$ dT/dt \propto \omega^{1.4} $ and $ \omega^{1.2} $ for pure water and salty ice, 
respectively. 
For salty water, we obtain $ dT/dt \propto \omega^{0.7} $. 
The proximity of the scaling laws for pure water and salty ice indicates 
that water molecules in salty ice under microwaves rotate as freely as those 
in pure water.

\begin{figure}
\centerline{\scalebox{0.50}{\includegraphics{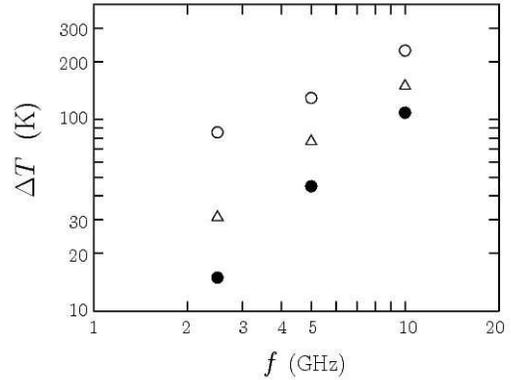}}}

\vspace*{-0.25cm}
\caption{
The dependence of heating rate on the microwave frequency. Open circles and 
triangels correspond to salty water and salty ice with 1 mol\% salinity, 
respectively, and filled circles correspond to pure water at 300 K.
}
\label{Fig.7}
\end{figure}

\begin{flushleft}
{\bf 4. Summary}
\end{flushleft}

By molecular dynamics simulation, we studied the heating process of salty ice, 
which we daily encounter when we melt frozen food in a microwave oven. 
We confirmed that the heating of salty ice is attributed to the weakening of 
the hydrogen-bonded network of water molecules in the presence of charged salt ions. 
In this process, the electrostatic effect was dominating over the short-range 
geometrical effect. 
The dependence of the heating rate on the microwave frequency was examined 
for salty ice and water, which again indicated that the internal structure of 
water molecules in salty ice is deformed. 
Thus, the water molecules in salty ice can rotate as freely as those in pure 
water under the microwave electric field.

\begin{flushleft}
{\bf Acknowledgments}
\end{flushleft}

The authors thank Dr. M. Matsumoto for fruitful discussions on the structure of ice. 
This work was supported by the Grant-in-Aid No.18070005 (FY2006-2010) from the 
Japan Ministry of Education, Science and Culture for Prime Area Research Project 
"Science and Technology of Microwave-Induced Thermally Non-Equilibrium Reaction Fields". 
Present computation was performed using our cluster machine consisting of 64-cpu 
Opteron 254 (2.8 GHz, single core) processors and high-speed InfiniBand interconnect.

\begin{flushleft}
{\bf References}
\end{flushleft}

\noindent
AMBER database and molecular dynamics code \\
\hspace*{0.1cm} ("Assisted Model Building with Energy Refinement"), \\
\hspace*{0.1cm} ver.10  (2008), at http://amber.scripps.edu/. \\
\noindent
Andersen, H.C. (1983) "Rattle: a velocity version of the \\
\hspace*{0.1cm} Shake algorithm for molecular dynamics calculations", \\
\hspace*{0.1cm} J.Comput.Phys. 52, 24-34. \\
\noindent
Berendsen, H., J.Grigela, and T.Straatsma (1987) "The \\
\hspace*{0.1cm} missing term in effective pair potential", J.Phys.Chem. \\
\hspace*{0.1cm} 91, 6269-6271. \\
\noindent
Buchner, R., J. Barthel and J. Stauber (1999) "The \\
\hspace*{0.1cm} dielectric relaxation of water between 0 C and 35 C", \\ 
\hspace*{0.1cm} Chem. Phys. Lett.  306, 57-63. \\
\noindent
Deserno, M. and Holm, C. (1998) "How to mesh up \\
\hspace*{0.1cm} Ewald sums. II. An accurate error estimate for the \\
\hspace*{0.1cm} particle-particle particle-mesh algorithm", J. Chem. \\
\hspace*{0.1cm} Phys. 109, 7678.\\
\noindent
"Dielectric Materials and Applications" (1954) edited by \\
\hspace*{0.1cm} R. von Hippel (MIT Press, Cambridge, USA).\\
\noindent
Eisenberg, D. and W. Kauzman (1969) "The Structure \\
\hspace*{0.1cm} and Properties of Water" (Oxford University Press, \\
\hspace*{0.1cm} London).\\
\noindent
English, N.J., and J.M.MacEloy (2003) "Hydrogen \\
\hspace*{0.1cm} bonding and molecular mobility in liquid water in \\
\hspace*{0.1cm} external electromagnetic fields", J. Chem. Phys. 119, \\
\hspace*{0.1cm} 11806-11813. \\
\noindent
English, N.J. (2006) "Molecular dynamics simulations \\
\hspace*{0.1cm} of microwave effects on water using different long-range \\
\hspace*{0.1cm} electrostatics methodologies", Mol.Phys. 104, 243-253, \\
\hspace*{0.1cm} and references therein. \\
\noindent
Hasted, J. B. (1972) "Liquid water: Dielectric \\
\hspace*{0.1cm} properties", in Water A comprehensive treatise, vol.1, \\
\hspace*{0.1cm} edited by F. Franks (Plenum Press, New York) 255-309.\\
\noindent
Hobbs, M.E., M.S. Jhon, and H. Eyring (1966) "The \\
\hspace*{0.1cm} dielectric constant of liquid water and various forms of \\
\hspace*{0.1cm} ice according to significant structure theory", Proc. \\
\hspace*{0.1cm} Nat. Acad. Sci., 56, 31-38. \\
\noindent
Matsumoto, M., S.Saito, and I.Ohmine (2002) \\
\hspace*{0.1cm} "Molecular dynamics simulation of the ice nucleation \\
\hspace*{0.1cm} and growth process leading to water freezing", Nature \\
\hspace*{0.1cm} 416, 409-413. \\
\noindent
Meissner, T. and F. J. Wentz (2004) "The complex \\
\hspace*{0.1cm} dielectric constant of pure and sea water from \\
\hspace*{0.1cm} microwave satellite observations", IEEE Trans. Geosci. \\
\hspace*{0.1cm} Remote Sensing 42, 1836-1849.\\
\noindent
Peelamedu, R.D., M. Fleming, D.K. Agrawal, and \\
\hspace*{0.1cm} R. Roy (2002) "Preparation of titanium nitride: \\
\hspace*{0.1cm} Microwave-induced carbothermal reaction of titanium \\
\hspace*{0.1cm} dioxide", J. Amer. Ceramic Soc., 85, 117-120.\\
\noindent
Roy, R., D. Agrawal, J. Cheng, and S. Gedevanishvili \\
\hspace*{0.1cm} (1999) "Full sintering of powdered-metal bodies in a \\
\hspace*{0.1cm} microwave field", Nature 399, 668-670.\\
\noindent
Sato, M. et al. "Science and Technology of Microwave- \\
\hspace*{0.1cm} Induced, Thermally Non-Equilibrium Reaction \\
\hspace*{0.1cm} Fields", Grant-in-Aid for Prime Area Research from \\
\hspace*{0.1cm} the Japan Ministry of Education, Science and Culture \\
\hspace*{0.1cm} (FY 2006-2010); see http://phonon.nifs.ac.jp/. \\
\noindent
Suzuki, M., Ignatenko, M., Yamashiro,M., Tanaka,M. \\
\hspace*{0.1cm} and Sato,M. (2008) "Numerical study @of microwave \\
\hspace*{0.1cm} heating of micrometer size metal particles", ISIJ \\
\hspace*{0.1cm} Intern'l 48, 681-684.\\
\noindent
Takei, I. (2007) "Dielectric relaxation of ice samples \\
\hspace*{0.1cm} grown from vapor-phase or liquid-phase water", Physics \\
\hspace*{0.1cm} and Chemistry of Ice, edited by W. Kuhs (Royal \\
\hspace*{0.1cm} Society of Chemistry, Cambridge) 577-584.\\
\noindent
Tanaka, M. and Sato, M. (2007) "Microwave heating of \\
\hspace*{0.1cm} water, ice and saline solution: Molecular dynamics \\
\hspace*{0.1cm} study",  J. Chem. Phys, 126, 034509, pp.1-9.\\
\noindent
Tanaka, M., H.Kono, and K.Maruyama (2008) \\
\hspace*{0.1cm} "Selective heating mechanism of magnetic metal oxides \\
\hspace*{0.1cm} by microwave magnetic field", Archive (arxiv) cond-mat \\
\hspace*{0.1cm} /0806-3055. \\ 
\noindent
Toukan, K. and A.Rahman (1985) "Molecular-dynamics \\
\hspace*{0.1cm} study of atomic motions in water", Phys. Rev. B31, \\
\hspace*{0.1cm} 2643-2648. \\
\noindent
Zhong, L., C. Xu, and C. Qiu (1988) "Anomalous high \\
\hspace*{0.1cm} permittivity in salty ice - a new dielectric \\
\hspace*{0.1cm} phenomenon", Proc. Properties and Applications of \\
\hspace*{0.1cm} Dielectric Materials, vol.1, 206-208.

\end{document}